\documentclass{article}
\usepackage{authblk}

\usepackage[utf8]{inputenc}
\usepackage[vertfit]{breakurl} 
\usepackage{amsmath}
\usepackage{amsthm}
\usepackage{amssymb}
\usepackage{color}
\usepackage{url}
\usepackage{enumitem}
\usepackage{outlines}
\usepackage{nicefrac}
\usepackage{soul}
\usepackage{pgfplots}
\usepackage{tabularx}
\pgfplotsset{width=10cm,compat=1.9}

\def\code#1{\texttt{#1}}

\title{NewsCompare - a novel application for detecting news influence in a country}
\author[1]{Cristian Pop}
\author[1,2]{Alexandru Popa}
\affil[1]{University of Bucharest}
\affil[2]{National Institute of Research and Development in Informatics}

\date{\today}

\begin{document}
\maketitle

\begin{abstract}

The concept of ``fake news'' has been referenced and thrown around in news reports so much in recent years that it has become a news topic in its own right. At its core, it poses a chilling question -- what do we do if our worldview is fundamentally wrong? Even if internally consistent, what if it does not match the real world? Are our beliefs justified, or could we become indoctrinated from living in a ``bubble''? If the latter is true, how could we even test the limits of said bubble from within its confines?

Without delving into advanced epistemology (i.e. fundamentally ``how do we know that we know?''), one of the obvious key requirements is avoiding a reliance on a single source of information. This is no easy task, considering the relative scarcity of truly independent news publishing companies, compared to say, massive conglomerates pushing a certain agenda under different publication names, or (possibly worse) state-sponsored propaganda. Fact-checkers going through articles individually can only do so much, before buckling under the never-ending deluge of information coming out, ranging from nearly-correct to completely fabricated. 

We propose a new method to augment this process, by speeding up and automating the more cumbersome and time-consuming tasks involved. Our application, NewsCompare takes any list of target websites as input (news-related in our use case, but otherwise not restricted), visits them in parallel and retrieves any text content found within. Web pages are subsequently compared to each other, and similarities are tentatively pointed out. These results can be manually verified in order to determine which websites tend to draw inspiration from one another. The data gathered on every intermediate step can be queried and analyzed separately, and most notably we already use the set of hyperlinks to and from the various websites we encounter to paint a sort of ``map'' of that particular slice of the web. This map can then be cross-referenced and further strengthen the conclusion that a particular grouping of sites with strong links to each other, and posting similar content, are likely to share the same allegiance.
We run our application on the Romanian news websites and we draw several interesting observations.
\end{abstract}

\section{Introduction}
\paragraph*{Motivation}

The topic of fake news is in the collective consciousness for some time now, due to its alleged impact on swaying public opinion on important issues, going so far as to potentially influence election results \cite{AllcottG2017} in some cases. We find entire articles devoted to studying their impact \cite{LazerM2018}, and methods of detection \cite{Conroy2015}. While some of the conclusions in these articles may be merely tentative, there are still some hard-to-dispute facts we can start using as a basis. For instance, we know that more than two thirds of Americans report getting at least some of their news on social media according to a Pew Research study \cite{ShearerE2017} from 2017. Worldwide, 48\% of people surveyed reported believing a fake news story was real before finding out it was fake, according to an Ipsos report \cite{Ipsos2018}. Interestingly, the same report finds that 63\% of people are confident in their own ability to identify fake news, while only 41\% are confident that the average person can do the same. Are people in general overly confident about themselves, or too cynical about others? Hard to say, but nevertheless an interesting idea to explore.

Enterprising research out there has already found insightful characteristics of fake news, with one paper going so far as to draw parallels between fake news and satire \cite{HorneB2017}. This could not be easily done without the appropriate technology to gather large quantities of data, and analyzing it in new and creative ways. Taken to its logical conclusion, such research could eventually lead to heuristic algorithms able to detect and filter out fake news, a monumental breakthrough in and of itself. While not being naive enough to ignore all the challenges (one could easily imagine the rise of an ``arms race'' between fake news manufacturers and detectors, akin to the current system of viruses and antiviruses), this is one idea that we have found immensely motivating in our quest to push the boundaries of what can currently be done.

Relatedly, examining how news sources disseminate their content, how this fits in to their respective ecosystem, and how they continuously adapt in order to keep up a working business model, are all intriguing subjects in their own right. We know from existing research that newspaper publishers are aggressively trying to expand into the digital realm, going as far as adopting a ``digital first'' approach, but the data shows they are still heavily reliant on print in terms of revenue \cite{MyllylahtiM2017}. Exclusively online news outlets on the other hand do not have the luxury of print to fall back on, so we expect them to make that much more of an effort in establishing a foothold in the online market to draw revenue from. This is actually supported by some of our findings, see Section \ref{sec:graph-analysis} for a specific example.

Since the topic of fake news is a complex one, it can hardly be expected to be tackled end-to-end over the course of a single article. More research is always welcome, and our understanding of it can only deepen in proportion with the number of researchers shining a spotlight towards it. Of course, any new research should ideally be done in a non-partisan fashion so that new studies can present objective conclusions, which are less likely to be dismissed offhand (especially by laypeople) in an increasingly polarised world \cite{CaryF2017}. That being said, it may be hard to even know how to begin tackling this issue, considering the sheer amount of data out there that needs to be collected, stored and whittled down into manageable chunks, to fit the scope of various investigations. As such, we want to do our part in reducing this barrier to entry, to build upon the works of others and at the same time provide a stepping stone for other people coming up with innovative research ideas that would otherwise be difficult to implement on account of technical challenges.

\paragraph*{Related work}

We find similar work already out there, albeit with slightly a different application and purpose. Of course, we are not the first to consider the potential of data analysis, and the usefulness of providing enthusiastic people with investigative acumen with tools they could put to good use. Gray et al \cite{GrayJ2012} offer a particularly accessible guide aimed at journalists wishing to take charge and initiate their own data-heavy investigations. There are also repositories \cite{InvestigativeDashboard} dedicated to collecting large troves of documents and other data sets, opening them up to be analyzed by interested parties. What we try to offer is a slightly ``meta'' spin, by enabling investigations into the supposed investigative outlets themselves. Keeping tabs on the behaviour of entities tasked with shaping public opinion, either deliberately or unwittingly, should arguably rank fairly high as far as research topics go.

The issue of scraping social media data is explored in some detail by Marres and Weltevrede \cite{MarresN2012}, who note that scraping is currently a prominent technique for the automated
collection of online data, promising to offer new opportunities for digital social
research. There is a fair amount of hype surrounding scraping as a herald of the coveted ``revolution'' in social research brought on by the advent of the Internet. What makes the technique special is allowing research to be done as an ongoing process, rather than a finished process. Of course, their application involved scraping just a handful of pages and charting very specific changes on said pages over time. Our application's current focus is a lot more generic, aiming to target a large number of distinct websites, and tries to avoid any kind of specialization that could prove restrictive for a general use case. Of course, future development can still be done to address various special cases with some minor tweaks.

The same article by Marres and Weltevrede \cite{MarresN2012} mentions a service used at the time, ScraperWiki \cite{ScraperWiki}, aiming to serve as a platform for developing and sharing scrapers. It has since been renamed to QuickCode, as it ``isn’t a wiki or just for scraping any more''. ScraperWiki is  mentioned a handful of times among the various works we have looked at in preparation for this article, but not so much since its rebranding as QuickCode. It is not entirely clear if the platform remains as accessible as it once was for the casual researcher at the time of writing. We could not find other similar platforms worth noting, therefore if web crawling/scraping research is indeed an underserved niche, our proposed solution should help plug that gap.

Other interesting research seeks to employ scraping to analysis with a more predictive application in mind. Lerman and Hogg \cite{LermanK2010} have tried come up with a model that is able to predict future news popularity starting from a data set acquired from scraping entries on a popular social media platform. Their work is greatly helped by the particular structure of their chosen platform (i.e. digg.com), where it is to pick up on early user voting results on new entries, extrapolating from there and estimating future popularity based. This should be easy to replicate on other sites with similar voting systems (e.g. reddit.com), but a great deal more creativity is required to do something similar on a more generic set of websites. That is, unless we can distill our set of target websites to include only ones with a very well defined set of characteristics, or choose some other metric to apply statistical modeling on and derive predictive benefit out of.

Yet another direction of research is sentiment analysis, as explored by Balahur and Steinberger \cite{BalahurA2009} specifically for the use case of news articles. They employ the freely accessible Europe Media Monitor (EMM) family of applications \cite{EMMNewsBrief}, which at the time was retrieving  between 80,000 and 100,000 articles per day in about 50 languages, scraping about 2,200
hand-selected online news sources and a few specialist websites (these numbers have increased in recent years). A fairly impressive data set, unless it happens that our target websites fall outside of these news sources, which is where our application fills in the gap by allowing any number of custom entries to scrape on a regular basis. We estimate that some fairly involved tweaks would be required to add a similar sort of functionality to the processing side of our application, but the website content as currently gathered by our scraper should already lend itself well to the task.

A more niche approach, coming from what looks like fledgling research from Vargiu and Urru \cite{VargiuE2013}, involves figuring out how to pick out the most relevant contextual ads, based on insight gleaned from from scraping existing web pages. This does not necessarily apply solely to news sites, but it does give us an idea of at least one of the lucrative directions this kind of research can develop into. The amount of automation already out there in the advertising world should give us pause for thought, however. A solid business model right now could prove to be overinflated and unsustainable in the long term. According to a 2014 study by Association of National Advertisers \cite{ANA2014}, bots now comprise an estimated 23\% of all online video ad viewers, and 10\% of all static display ads. Rushkoff presents an eloquent, yet grim (and possibly somewhat alarmist) view in his book \cite{RushkoffD2016} on the topic:

\begin{quote}
\textit{Consider the irony: malware robots watch ads, monitored by automated tracking software that tailors each advertising message to suit the malbots' automated habits, in a human-free feedback loop of ever-narrowing ``personalization''. Nothing of value is created, but billions of dollars are made.}
\end{quote}

With that in mind, we should be far more interested in creating something of value, rather than chasing ephemeral gains.

\paragraph*{Our results}

What we try to add to the existing body of work is effectively a new solution in the form of a fast, efficient, mostly automated application able to gather vast amounts of information about websites, in as generic a form as possible. Our aim is to have an information dump that is easily to compile, and greatly simplifies the work of future researchers who need large sample sizes to interpret and derive conclusions from, according to various specific use cases. Some of these use case ideas have already been at least tentatively explored in articles mentioned in this introduction. We are confident that a good deal of research endeavors would have benefited from the kind of data dump we can now provide, and yet more research can benefit from it going forward.

We also put the application to the test on an individual use case to start with (i.e. Romanian news websites), to at least overcome the most glaringly obvious issues and challenges before releasing to the general public. A good deal of effort has been made in ensuring the application has more than just a niche appeal about it, and that it can be run reliably for long stretches without much manual interference. However, we also expect (and welcome) any constructive criticism and bug reports that get us closer to a flawless product. Despite not coming from a sociology background, we try our hand at interpreting the results we get from our use case, at least to the extent that we are aware of what characteristics to look for (see Section \ref{sec:use-case} for more details).

In the process of developing the app, some of the biggest hurdles that had to be handled were caused by the flaky and unpredictable nature of web content in general. By far, the element of human error involved in setting up websites seems to be the biggest source of issues with setting this sort of automated solution. Simple typos can lead to cascading failures (sometimes in spectacular fashion) when improperly interpreted by our heuristic algorithms. These failures are typically only obvious when they get to the point where they manifest among a noticeable segment of our result set. As such, there is a wide range of special case handling baked into our application code. While probably not fully exhaustive, we can reasonably expect that scenarios that are yet to be discovered should not have a statistically significant impact on results.

\section{Overview of the application}

The back-end runs as an executable JAR file, so the machine running it needs to have Java runtime installed (version 8 or later). We also need to set up a PostgreSQL database for it to use, which can be done by following the steps listed on the GitHub project page readme file \cite{GitHubNewsCompareBackend}. Once started, it will start automatically start crawling any sites listed in its database (this will be empty to start with). While crawling, it scans for new links to visit, and download the text content from every website visited to a local folder, where it will be indexed and processed to find similarities. More technically-inclined users will be able to connect to the back-end database directly to view real-time changes and make any low-level tweaks where it makes sense to do so. 

The front-end is a Javascript-based single-page app (SPA) serving a number of key functionalities:
\begin{itemize}[noitemsep]
  \item Listing all websites discovered by the web crawler
  \item Allowing specific websites to be toggled as special interest, Romanian news websites for our use case, causing them to be queried more often for snapshots (at minimum every 1 hour)
  \item Drawing a graph to visualize links to and from our special interest websites, allowing nodes to be added or removed in order to minimize clutter
  \item Listing instances where similar text content was detected on different websites
  \item Displaying various statistics about the web crawler's activity
\end{itemize}

Note that the front-end can only be accessed while the back-end is running. A fully featured implementation for our use case is available online at \url{http://www.newscompare.tech} \cite{NewsCompareTech} for demonstration purposes.

\section{Technical details}
\label{sec:technical}

All the code written for the application is freely available on GitHub \cite{GitHubNewsCompareBackend,GitHubNewsCompareFrontend} for anyone to examine or make use of, either as-is or by building upon it to suit some other purpose. In its current form, it should be accessible to most developers (particularly coming from a Java background), by following the instructions listed in the ``Readme'' files. For reference, the desktop machine used for all development and testing work is running a 64-bit octa-core (16-thread) CPU  with 64 GB of RAM, and an NVMe solid-state drive for storage. The application is not particularly memory-intensive, but due to its multithreaded nature it does benefit significantly from multiple cores and high speed CPUs. Depending on the number of websites targeted and the frequency of snapshots taken, available storage could start to become a concern. For just over 100 websites each snapshot folder seems to add up around 1 gigabyte, including website text content and generated indexes.

Note that the back-end project is the most important component, and will be referred to interchangeably as ``NewsCompare'' or ``the application'' throughout this article. It can be picked up from scratch, and used on its own just by tweaking configuration values and keeping an eye on logging output. The front-end is effectively just a convenient way of interacting with the data set and provides some visualization of the application results. A prebuilt version of the back-end component is available on GitHub \cite{GitHubNewsCompareBackendJARFile}, with all the default settings we used throughout our testing configured at compile time.

In Subsection \ref{sec:web-crawling} we try to give a comprehensive run through the complexities of developing a web crawling solution nearly from scratch, which may prove helpful to anyone interested in rolling their own implementation. Subsection \ref{sec:text-comparison} similarly deals with how we set up an inverted index \cite{BlackP2008InvertedIndex} for our set of text documents in order to perform fast searches and comparisons between them. This whole section should be a good starting point for anyone trying to better understand our publicly available code, to either modify or improve it. We try to note various issues and improvements  already tackled, and also lay out some potential quality-of-life improvements for the future.

\subsection{Web crawling}
\label{sec:web-crawling}

According to the comprehensive primer by Olston and Najork \cite{OlstonN2010}, a web crawler (also known as a robot or a spider) is a system for the bulk downloading of web pages. Web crawlers are used for a variety of purposes, and the one we are most interested in here is an application of data mining, where we analyze web pages for statistical properties, and try to perform various data analytics. The web crawler starts off with a list of URLs to visit, otherwise known as \textit{seeds}. This list can be quite small to begin with, as we expect it to grow exponentially. As the crawler visits these URLs, it identifies all the hyperlinks in the page and adds them to the list of URLs to visit, known as the \textit{URL frontier} in some publications \cite{ManningC2008}. URLs from the frontier are recursively visited according to a set of policies. If the crawler is performing archiving of websites, as we do, it copies and saves the information contained within as it goes.

In our particular case, we try to visit every hyperlink at least once, but place a much higher emphasis on a manually curated list of websites where we want frequent snapshots saved. How often we are able to take these snapshots largely depends on how quickly we can run through this list on every iteration. As long as we keep it relatively short, and only visit a small set of websites on every iteration, we can afford to schedule our crawler to run fairly often. This in turn allows taking frequent snapshots, which are useful for record-keeping or auditing purposes. The main graphs and reports that we generate should typically be based on the most recent snapshot, unless a specific comparison between snapshots is otherwise required.

One of the immediately useful side effects of web crawling is that we automatically get to compile a list of directional links between the websites we start off with, and the ones discovered along the way. This allows us to effectively map a limited section of the visible web, and visualize it as a directed graph, with the websites serving as nodes and links as directed edges. This can serve as a basic sanity check on whether our results look valid and useful, but can also lead to basic conclusions in their own right (provided we have some interest in web architecture to begin with). Having readily available access to basic graph details, like node degree and connectivity allows us to see how our results line up with existing research, and potentially put it to the test. See section \ref{sec:use-case} for some actual examples of insight derived from our particular set of target websites.

\subsubsection{Challenges}
\label{sec:challenges}

\textbf{Successive requests} to the same server can lead to getting blacklisted or banned if the time between requests is too short. Should this occur on some websites (and slip by unnoticed), it could potentially skew our result set. Olston and Najork put it very succinctly in their survey on the science and practice of web crawling \cite{OlstonN2010}:
\begin{quote}
  \textit{Crawlers should be “good citizens” of the web, i.e., not impose too much of a burden on the web sites they crawl. In fact, without the right safety mechanisms a high-throughput crawler can inadvertently carry out a denial-of-service attack.}
\end{quote}
A naive implementation of a web crawler might overlook this (or a malicious actor could ignore it entirely), but it stands to reason that most servers will rightfully seek to defend themselves against perceived acts of aggression, at least to the extent of limiting damage and maintaining high uptime. Any behavior that would not realistically be carried out by a human could raise red flags, causing web servers to start dropping requests. Based on our own empirical observation, imposing a mandatory delay of 100-200ms between successive requests seems to yield good results.

\textbf{URL normalization}~\cite{rfc3986} is an important requirement, at least to the degree where we are satisfied that it covers our target websites properly. Any shortcomings in this area could lead to visiting semantically equivalent URLs, resulting in wasted effort and potential over-representation in our result set.

From the existing research, we adapt a few interesting ideas from a proposed algorithm \cite{Bar-YossefZK2009} for a systematic and robust method of URL normalization, however in the final implementation we rely largely on simple string manipulation using regular expressions. Some of the more notable steps we employ are:
\begin{itemize}[noitemsep]
  \item Converting the scheme and host to lower case
  \item Removing the default port (e.g. 80 for \code{http})
  \item Removing the fragment (\code{\#}) and query (\code{?}) components of the URL
  \item Removing the protocol component (\code{http://} or \code{https://})
  \item Removing \code{www} as the first domain label (where it exists)
\end{itemize}

\textbf{Filtering out} non-web content helps keep our data set smaller and more focused. We employ several basic methods here, based on string pattern matching in website URLs. The links we filter out here are not web pages, so we know from the start they are unlikely to provide useful information in keeping the crawling process going:
\begin{itemize}[noitemsep]
  \item Filtering by file extension, e.g. links ending in ``.jpg'', ``.doc'', ``.avi'', ``.mp3'' etc.
  \item Filtering non-HTTP(S) protocols, e.g. links starting with ``mailto://'', ``skype://'', ``whatsapp://'' etc.
  \item Filtering specific string formats, e.g. links formatted in a way consistent with phone/fax numbers, or email addresses.
\end{itemize}

\textbf{``Crawler traps''} can be a significant time drain if unnoticed for extended periods. As Olsten and Najork mention~\cite{OlstonN2010}, there exist ``websites that populate a large, possibly infinite URL space on that site with mechanically generated content''. The example they give is that of a web-based calendaring tool, where each month has its own page and a hyperlink to the next and previous months. 

For our given set of websites, the biggest danger we notice is that of websites linking to various external indexing services. For instance, a website could link to its own entry on archive.org, which sends our crawler down an unfeasibly long chain of links that do not really improve our result set if followed. We are not directly interested in travelling the entire breadth of other existing indexing services or aggregators, we have started to maintain a list of exclusions for the crawler to avoid. See \ref{sec:improvements} for an idea on improving this process.

\textbf{Thread-safe} methods for reading and writing to data storage, in the context of using a single, traditional SQL database for data storage. In this case, using a transaction isolation level of ``repeatable read'' in PostgreSQL \cite{postgres-transaction-isolation} appears to be enough to ensure the data integrity with only a moderate slowdown.

\subsubsection{Optimization}
\label{sec:crawling-optimization}
\textbf{Parallelization} is something generally well-suited to web crawling activities. Intuitively we should be able to fetch content for most websites independent of each other, so the work can be done on separate threads. Some experimentation may be required with the number of threads assigned to each individual task in order to achieve this result.

Java's parallel streams \cite{JavaParallelism} are a handy tool for quickly implementing parallel processing. Where other, more low-level, solutions would require us to handle the dividing of a problem into subproblems, then combining the results of the solutions ourselves, the Java runtime does this largely automatically. While it does not automatically guarantee operations perform faster (in some cases, quite the opposite, due to overhead), it makes it quite easy to make small code tweaks and find an optimal solution through trial and error. If we already follow the functional programming paradigm, it could mean something as simple as replacing calls to \code{stream()} with \code{parallelStream()}, configuring the thread pool size, and running performance tests. See Table \ref{table:ParallelStreamTable} for the results of a test run comparing the effect of various configuration settings on the same sample set of 334 web pages.

\begin{table}[ht!]
\centering
\begin{tabular}{||c c c||} 
 \hline
 No. of threads & Run time (s) & Average pages/s \\ [0.5ex] 
 \hline\hline
 1 & 665.3 & 0.50 \\
 5 & 239.4 & 1.39 \\
 10 & 130.2 & 2.56 \\
 20 & 74.5 & 4.48 \\
 50 & 27.2 & 12.27 \\
 100 & 25.6 & 13.04 \\ 
 200 & 33.4 & 10.00 \\ [0.5ex] 
 \hline
\end{tabular}
\caption{Effects of multithreading on web crawling speed (334 pages sample size)}
\label{table:ParallelStreamTable}
\end{table}

When crawling deeper than surface-level (i.e. more than just the website home page), we need to be cautious about our parallel tasks from inadvertently making too many simultaneous requests to the same server, as we mention in Subsection \ref{sec:challenges}. Our approach here is divide our entire set of target web pages we wish to crawl during a regular run (typically around 10-20,000 for our 334 target websites) into ``slices'', with each slice containing at most one page from the same parent domain. These slices can be visited entirely sequentially, which is nice and safe (but also slow), or we can try to find a way to have them run in a partially overlapping fashion, which would be more optimal but also place us at a slightly higher risk of getting blacklisted for excessive requests. In our particular case, we settle on configuring each thread to run at a slightly different delay (in 100 ms increments), which does not seem to impact the rate of successful requests, and the speed improvement is more than threefold for our use case.

\begin{table}[ht!]
\centering
\begin{tabular}{||c c c c c c||} 
 \hline
 Mode & Pages & Successes & Success rate & Time (s) & Avg. pages/s \\ [0.5ex] 
 \hline\hline
 Sequential & 10434 & 9533 & 91.36\% & 788.95 & 13.22 \\
 Overlapping & 9163 & 8296 & 90.53\% & 212.49 & 43.12 \\ [0.5ex] 
 \hline
\end{tabular}
\caption{Effects of fetching webpages slices sequentially vs. overlapping}
\label{table:ParallelSlicesTable}
\end{table}

\textbf{Separating tasks}, i.e. having the raw data retrieval separate from the actual processing, should increase overall throughput. Since network speed and latency can vary wildly by server, hitting a particularly slow website might otherwise bottleneck the entire process. 

Depending on the particular use case, we can wait until our targeted websites are fully retrieved before starting the processing, or we can run the two tasks roughly in parallel, with the processing lagging slightly behind. Some factors influencing this decision include whether we want to extract useful information from partial results, or if we think we could squeeze some extra performance and have the CPU cycles to spare for it (i.e. if fetching websites is not already keeping us at 100\% load, or close to it).

\textbf{Batch processing} is one of the less obvious points, since in typical work loads with small sample sizes the performance impact is negligible. But something like building a large array of objects that are saved in a single call to the database, instead of saving each one individually gives us a massive speed advantage, particularly in the context of multiple threads seeking concurrent database access.

\textbf{Timeout periods} should be configured to a sensible value, to prevent having threads locked up in useless waiting periods for more than is absolutely necessary. This value can be arrived at through trial and error, by keeping track of the number of successful server responses over multiple trial runs. We expect this number to increase along with the timeout period, but we should see the improvement rate drop off sharply after a certain point. It is precisely this point of diminishing returns that gives us the best trade-off between results and performance.

We present the results of several web crawler test runs, each with different timeout values, where we target 334 websites and request 20 distinct pages from each of them.

\begin{table}[ht!]
\centering
\begin{tabular}{||c c c c c||} 
  \hline
  Timeout (s) & Run time (s) & Requests & Successes & Percentage \\ [0.5ex] 
  \hline\hline
  0.001 & 504.0 & 11420 & 9766 & 85.51\% \\ 
  1 & 530.4 & 11420 & 9732 & 85.21\% \\
  5 & 622.9 & 11420 & 9760 & 85.46\% \\
  10 & 740.4 & 11420 & 9773 & 85.57\% \\ [0.5ex] 
  \hline
\end{tabular}
\caption{Effects of timeout value on crawler speed and success rate}
\label{table:TimeoutSuccessTable}
\end{table}

As we expect, the largest timeout values correlate with the largest number successful requests, but the improvement rate is marginal at best. The difference in percentage points is so insignificant enough that it may be explained away by random chance (or possibly, random background CPU usage on the test machine at the time). As such, we are comfortable in reducing the timeout to the bare minimum for our purposes.

\subsubsection{Future improvements}
\label{sec:improvements}
\textbf{Automating ``crawler trap'' detection} is a good starting step for improving the robustness of the application, allowing it to run independently with greater confidence. Since there is no real way to predict how often this kind of issue can surface, the way we mitigate it currently is by keeping an eye on the application's log output on a regular basis. We need to manually add any newly discovered ``trap'' to our list of excluded domains, so this is somewhat time-consuming. As soon as we find the trade-off acceptable, we can look at implementing a heuristic algorithm limiting the crawler's traversal of a particular domain past a specific threshold, making the entire process more automated.

\textbf{Smarter timeouts} would improve general crawling speed, especially over long stretches of time, but the impact can vary between marginal and significant depending on the set of websited targeted. By keeping track of each website's recently failed requests, we can place servers that seem to be unreachable (temporarily or otherwise) on cooldown, querying them less often and reducing the amount of time wasted waiting on timeouts overall. 

The cooldown value can be set to an arbitrary value to start with, but ideally should be arrived at after some experimentation. We do not want to unwittingly restrict certain websites from our data set too harshly and risk skewing our conclusions. However, since any websites affected by this optimization are unresponsive to begin with, the risk of this should be fairly low.

\textbf{Database replication} adds a fair degree of complexity to the entire architecture, but at the same time provides a small-to-moderate boost in performance, by separating the application's responsibilities among multiple databases hosted on potentially multiple servers (or virtual machines). We expose a good number of API endpoints, some for displaying various statistics on the crawler's progress and results, others to provide a better visualization on our data set (or sections thereof). The SQL queries involved in retrieving this data take up to several seconds to run in some cases, largely due to the volume of data involved, and this constitutes extra load on our current ``single point of failure'' database.

PostgreSQL provides a very powerful solution in this regard \cite{postgres-streaming-replication}, allowing us to do near real-time streaming replication of data from our master database to a standby one. The former can keep handling all the ``heavy lifting'' required by the web crawler, while the latter is used as a read-only source for reporting purposes. We also get the added bonus that the standby database can be automatically promoted at any time to master status, should the original master suffer an unrecoverable failure, significantly improving uptime and reliability. We do not include an implementation of database replication in the current version of the app, as it would further complicate the setup process for anyone seeking to reproduce (or build upon) our findings. It is however worth mentioning, in the interest of laying out the various pros and cons for interested parties.

\subsection{Text comparison}
\label{sec:text-comparison}

Once we get the process of acquiring a large data set of website content out of the way, the next step is to do more in-depth processing and extract more valuable information out of it. What we want to do is implement a kind of plagiarism detector to point out the more glaring similarities between articles on different websites. From the outset, it is clear this can turn into a time and resource-intensive task, and we need to be somewhat clever in order to avoid exponential complexity spiraling out of control and rendering the whole thing unfeasible. 

\subsubsection{Simple approaches}
Without even delving into algorithms, it should be make intuitive sense that most naive implementations  would require too much processing to allow it to scale well, and there is at least some minimal research required to avoid wasting much time reinventing the wheel. For instance, a brute-force method of comparing 100 pieces of text one by one would require 4950 separate comparisons after a quick calculation (\nicefrac{n(n-1)}{2} where n = 100). Any optimization we implement along the way to reduce the number of comparisons performed can have a significant impact on the overall time. The particular algorithm we implement for performing the comparison is also crucial, to the extent that we can find one to process chunks of website text at least as fast as they are coming in from the web crawler side.

Computing a kind of string similarity coefficient based on Levenshtein distance \cite{BlackP2008Levenshtein} (i.e. finding the smallest number of insertions, deletions, and substitutions required to change one string or tree into another) potentially gets us the results we are interested in, but is still very much a brute force approach. The most obvious shortcoming is that we are effectively doing the same work over and over by processing each string from scratch on every comparison. The first big improvement would be to introduce an initial, preparatory step of distilling strings into their base components for easier comparison later on.

\subsubsection{More advanced approaches}
Donald Knuth gives a very well-written primer in his famous book \textit{The Art of Computer Programming} \cite{KnuthD1997} on how inverted indexes are used to set up fast searching through text strings. To put it succinctly, we set up our index by making a list of unique terms in each individual block of text, and keep track of where the term is located within the text. From here, we can boil down every word to its most basic form (e.g. plural to singular, conjugated verb to infinitive form etc.) to reduce the size of our list of terms while improving representation. Additionally, we can filter out so-called ``stop words'', which are the most frequent and almost useless words (e.g. ``a'', ``I'', ``the'' for English), further lowering the noise in our search results.

Luckily, we are able to avoid much of the complexity of implementing our own inverted index solution by co-opting the open-source project Apache Lucene \cite{ApacheLucene} into the application. It comes with a wide array of language analyzers (including Romanian), making it suitable both for our particular use case and improving the odds of our application becoming useful as a generic tool for future researchers. By making good use of Lucene's ``more like this'' functionality \cite{ApacheLuceneMLT}, we can avoid making an inordinate amount of one-to-one comparisons between items in our data set. This largely mitigates one of the concerns stated earlier, and means the number of comparisons we do (as well as the time taken for each comparison) should scale linearly rather than exponentially.

At this point we are able to perform the indexing and comparison steps at a manageable pace, something in the order of minutes instead of days for around ten thousand text files. However, we still have significant noise in our result set, so we need to further refine our algorithms. To this end, Abid et al \cite{AbidM2017} suggest n-grams, i.e. sequences of words of length n, are a much better choice than single words for indexing and searching. Indeed, we observe a much tighter result set after switching to tri-grams, and the set itself is small enough to be discernible by a quick skim through (no longer requiring us to scour through millions of resulting combinations).

\subsubsection{Challenges}
Most challenges in this area stem from the fact that we are attempting to adapt a number of rough, heuristic algorithms to make sense of fairly nuanced text generated by humans (i.e. an extremely limited application of natural language processing). We want it to be useful, so the signal to noise ratio needs to be high, without excluding any useful results and reduce our overall accuracy. For instance, some of the conclusions coming from the app may be accurate (two pieces are text are very similar), but effectively useless at the same time (e.g. copied and pasted cookie policies, privacy policies, GDPR statements etc.). Conversely, two sources may be very similar content-wise, but the individual website's HTML structure could make it difficult to pick out particularly relevant blocks of text, causing it to slip under our radar.

\textbf{Website architectures} can be quite varied, and we want to keep any assumptions about particular approaches in this field at a minimum, so that the application ca be as generic as possible. In particular, subdomains can be somewhat tricky to deal with, we need to remember at all times to consolidate results belonging to the same top domain as a single source. After all, our stated purpose is to find similarities between wholly distinct websites, to point out the spread of content, and we do not concern ourselves with reused content between different sections of the same website. This consolidation step goes a long way towards improving our signal-to-noise ratio and making the more interesting results shine through.

\textbf{Relevant content} is sometimes hard to discern from surrounding context. Looking at any given news website, there is a lot of content displayed on page, but there is often surprisingly little space alotted to the actual content, i.e. the news article itself. The sidebars are typically reserved for internal/external links, advertising, and various widgets seemingly designed to provide some kind of use to the reader. We can discern that many design patterns favor drawing the user's attention, keeping them engaged and encouraging repeated visits, even when it might come into conflict with the main stated purpose of the site. While humans can quickly learn to intuitively pick up on useful content, automating this kind of processing into our algorithms can be quite tricky and time-consuming. In particular, the rise of interstitial advertising, and a general tendency to break up news into fragments and sprinkle  vaguely related content between them needs to be accounted for. We will not go into whether or not an entire article is effectively an advertisement, as that falls somewhere outside the scope of the current research.

\subsubsection{Optimization}
\label{sec:text-comparison-optimization}
\textbf{Parallelization} is already mentioned in \ref{sec:crawling-optimization} with regard to how it dramatically improves web crawling performance. The same rules apply here, even though we may not be able to find the same number of truly independent tasks that can be run in parallel. We hit a plateau of diminishing returns fairly quickly, but the performance gains are still worth pursuing as long as they are not too time-draining or significantly impact the readability or maintainability of the resulting code. We sit at a comfortable level of throughput right from the start, in no small part thanks to inherent optimizations present in the software library we employ \cite{ApacheLucene}.

\subsection{Technologies used}
\label{sec:tech-used}

We aim to avoid using any proprietary or license-based software, so that all of our code can remain public. We are grateful to the open source community for the multitude of varied and powerful tools at our disposal, and we can at least state that we do not feel hamstrung by our decision. An honorable mention should be made to \textbf{Apache Nutch} \cite{ApacheNutch}, a fully featured web crawling solution that could help future tech-minded people to co-opt web crawling into their projects. We do not make use of Nutch in our case, mainly because we wanted to have tighter control over the crawler's behavior, and were comfortable enough in rolling our own lower-level implementation.

\subsubsection{Back-end}

We use \textbf{Spring Boot} \cite{SpringBoot} to quickly and easily get a RESTful web service \cite{PautassoC2008} up and running using Java, but with minimum boilerplate and configuration out-of-the-box. It ties in well with \textbf{PostgreSQL} \cite{PostgreSQL}, which is used for mostly for persistence, but also storage to some degree. We need to save a limited amount of data from the websites explored by the web crawler to our database, some of which is used to inform future crawling iterations.
\textbf{Hibernate} \cite{Hibernate} makes it easier to perform the mapping between our Java classes and database tables, while \textbf{Flyway} \cite{Flyway} allows us to create our database structure in incremental migrations that can be easily replayed on a new machine when setting it up from scratch.

The crawler component uses \textbf{jsoup} \cite{jsoup} to create all of its network connections and also parse the resulting HTML pages using methods that allow for familiar CSS-like selectors. We also make local text dumps of the bulk of website contents, which are afterwards picked up by our implementation of \textbf{Apache Lucene} \cite{ApacheLucene}, creating indexes for quick text searches and comparisons.

\subsubsection{Front-end}

We use \textbf{Knockout} to build a simple yet dynamic JavaScript interface that pulls data from our application's endpoints and displays them in a more user-friendly fashion. The graph page uses an implementation of \textbf{vis.js} to help visualise website data as an interactive graph, again using data pulled from the back-end. \textbf{Webpack} is used to create a browser-friendly bundle of our own JavaScript source files, together with any node packages we use, as well as any other assets (e.g. CSS files).

\section{Use case: an analysis on Romanian news websites}
\label{sec:use-case}

To test our application, we define a particular use case by restricting the web crawling and analysis to a limited geographical area. We have made this choice largely in the interest of a fast turnaround time, to be able to make quick, experimental changes to our algorithms and study their impact immediately. We avoid making hardcoded assumptions, so that any tools we use can be repurposed with a different scope in mind, large or small.

Romania is actually an interesting choice in this respect, boasting a number of surprising, confusing, or ultimately even paradoxical characteristics. We consider the country to be in a rather unique position with regard to the relationship of Romanians with their fulfillment of basic needs and wants, one of which being news and media consumption. For an unassuming mid-sized country on the geographical fringe of the European Union, it boasts the highest average peak internet speeds in the European region, and is ranked at number 10 worldwide, according to Akamai's 2016 report \cite{Akamai2016}. Coupled with the generally affordable access plans (both wired and wireless), it is no wonder that adoption rates are on a continuous upwards trend, reaching 81\% in 2018 among the 16-74 year old population, according to Eurostat data \cite{EurostatInternet2018}. While still slightly below EU average (89\% in the same Eurostat data set), if we extrapolate from existing trends we could speculate that the adoption rate should reach the EU average in due course.

A recent report by Reuters Institute \cite{DigitalNewsReport2018} states that 88\% of Romanians get their news online, 82\% from TV, 67\% from social media and just 18\% from print. We can therefore expect that online news websites hold significant sway in shaping public opinion, considering the significant section of the populace relying on them. A similar report from the year before \cite{DigitalNewsReport2017} finds that ``the Romanian news environment is defined by intense competition for television and online audiences, sustained by understaffed newsrooms that struggle for financial survival''.

\subsection{Graph analysis}
\label{sec:graph-analysis}

We start our study by directing the application towards the top news websites by monthly popularity \cite{TraficRo}. From there, we get a record of all links encountered, which can be later visualized as the edges on a directed graph. What we end up with is a fairly large grouping of websites, centered around a smaller core of websites that we are actively interested in (i.e. Romanian-language news websites). The grouping itself is interesting, if we lay out a visual representation of the graph we can see that the links are not formed at random, and are more heavily weighted towards some websites. On one end, there are very few sites with a great number of links, and on the other end many sites with a very small number of links. A linear plot makes it harder to notice this fact, so we need to create a log-log plot to make it stand out more in our distribution. Our plots seem to line up with conclusions from existing research targeting internet topology \cite{FaloutsosM1999}, claiming that we should expect to see a surprisingly simple set of power-laws that describe concisely skewed distributions of graph properties such as the node outdegree and indegree.

For instance, we can see a distinctly non-random pattern if we look at the entire set of Romanian-language websites (not just news) that our crawler has visited at least once. The plots below display data points for roughly 65,000 Romanian websites found in this manner.

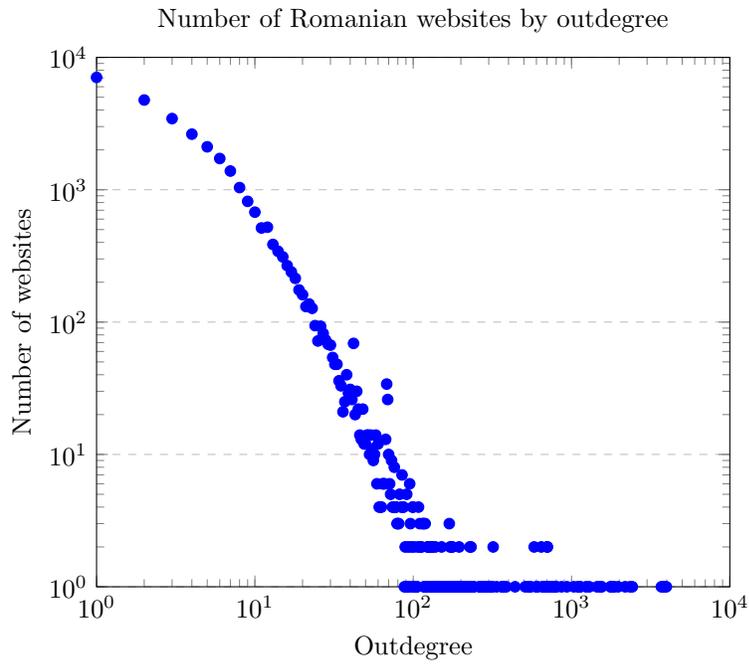
\begin{tikzpicture}
\begin{loglogaxis}[
    title={Number of Romanian websites by outdegree},
    xlabel={Outdegree},
    ylabel={Number of websites},
    xmin=1, xmax=10000,
    ymin=1, ymax=10000,
    xtick={1,10,100,1000,10000},
    ytick={1,10,100,1000,10000},
    legend pos=north west,
    ymajorgrids=true,
    grid style=dashed,
]
\addplot[ color=blue, only marks ]
    coordinates {
    (1,7058)(2,4761)(3,3448)(4,2629)(5,2110)(6,1721)(7,1384)(8,1038)(9,817)(10,677)(11,514)(12,521)(13,386)(14,343)(15,311)(16,267)(17,239)(18,214)(19,175)(20,161)(21,131)(22,137)(23,127)(24,94)(25,72)(26,93)(27,82)(28,73)(29,68)(30,67)(31,54)(32,48)(33,48)(34,36)(35,33)(36,21)(37,25)(38,40)(39,29)(40,31)(41,26)(42,69)(43,20)(44,30)(45,22)(46,14)(47,13)(48,22)(49,12)(50,12)(51,14)(52,14)(53,10)(54,14)(55,11)(56,9)(57,10)(58,14)(59,6)(60,12)(61,4)(62,4)(63,4)(64,6)(65,6)(66,6)(67,13)(68,34)(69,26)(70,10)(71,6)(72,5)(73,9)(74,4)(75,4)(76,8)(77,4)(78,4)(79,3)(80,3)(81,3)(82,5)(84,4)(85,7)(86,4)(87,4)(88,1)(89,2)(90,1)(91,5)(92,1)(93,2)(94,1)(95,6)(96,3)(97,2)(98,1)(99,4)(100,4)(101,2)(102,1)(103,1)(104,1)(105,1)(106,1)(107,2)(108,4)(110,3)(111,2)(113,2)(115,3)(116,3)(117,1)(118,1)(119,3)(121,1)(123,2)(124,1)(125,2)(126,2)(129,1)(130,2)(131,1)(132,1)(133,2)(134,1)(136,1)(138,2)(139,1)(140,1)(142,1)(146,1)(150,1)(151,2)(156,1)(158,1)(165,1)(167,1)(168,1)(169,3)(171,2)(172,1)(173,1)(175,2)(177,2)(178,1)(185,1)(186,1)(190,1)(195,2)(198,1)(202,1)(203,1)(209,1)(215,1)(221,1)(222,1)(223,1)(225,1)(226,1)(228,2)(229,1)(232,2)(240,1)(242,1)(261,1)(265,1)(267,1)(269,1)(276,1)(277,1)(281,1)(286,1)(295,1)(299,1)(306,1)(313,1)(317,1)(320,2)(337,1)(364,1)(365,1)(369,1)(371,1)(374,1)(387,1)(440,1)(510,1)(534,1)(543,1)(553,1)(581,2)(595,1)(613,1)(644,2)(656,1)(662,1)(689,1)(702,2)(704,2)(708,1)(714,1)(717,1)(718,1)(720,1)(723,1)(729,1)(732,1)(742,1)(768,1)(777,1)(782,1)(807,1)(865,1)(935,1)(1036,1)(1107,1)(1111,1)(1116,1)(1128,1)(1135,1)(1235,1)(1296,1)(1451,1)(1519,1)(1553,1)(1763,1)(1781,1)(1842,1)(1852,1)(1968,1)(2182,1)(2339,1)(2421,1)(3694,1)(3800,1)(3838,1)(3971,1)(14875,1)(16938,1)(38198,1)
    };
    \legend{}
\end{loglogaxis}
\end{tikzpicture}

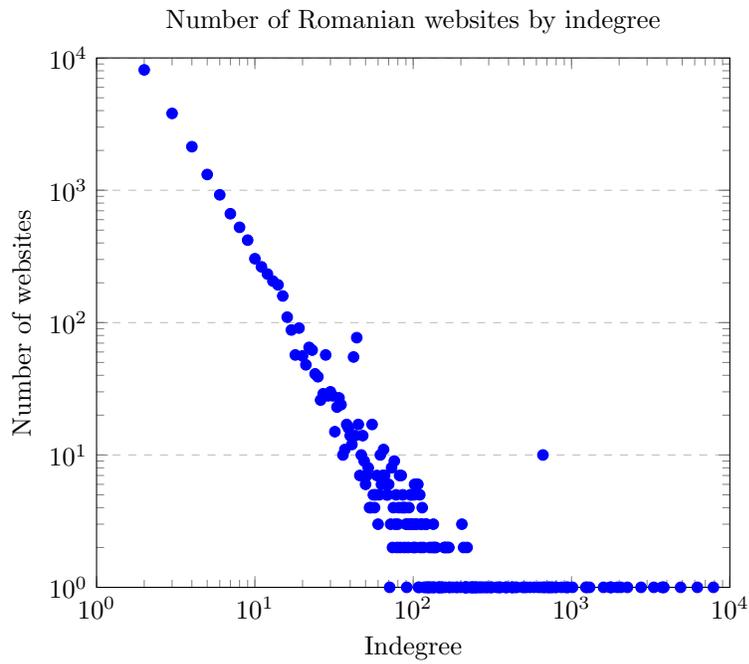
\vspace{0.25cm}

\begin{tikzpicture}
\begin{loglogaxis}[
    title={Number of Romanian websites by indegree},
    xlabel={Indegree},
    ylabel={Number of websites},
    xmin=1, xmax=10000,
    ymin=1, ymax=10000,
    xtick={1,10,100,1000,10000},
    ytick={1,10,100,1000,10000},
    legend pos=north west,
    ymajorgrids=true,
    grid style=dashed,
]
\addplot[ color=blue, only marks ]
    coordinates {
    (1,21830)(2,8123)(3,3808)(4,2135)(5,1316)(6,924)(7,665)(8,525)(9,420)(10,304)(11,264)(12,233)(13,206)(14,193)(15,159)(16,110)(17,88)(18,57)(19,91)(20,56)(21,48)(22,65)(23,62)(24,41)(25,39)(26,26)(27,29)(28,57)(29,28)(30,30)(31,28)(32,15)(33,23)(34,27)(35,24)(36,10)(37,11)(38,17)(39,16)(40,14)(41,12)(42,55)(43,14)(44,77)(45,17)(46,7)(47,10)(48,14)(49,9)(50,6)(51,7)(52,8)(53,4)(54,4)(55,17)(56,5)(57,4)(58,5)(59,7)(60,3)(61,5)(62,10)(63,6)(64,7)(65,11)(66,7)(67,6)(68,5)(69,5)(70,6)(71,1)(72,3)(73,8)(74,2)(75,4)(76,9)(77,3)(78,5)(79,2)(80,3)(81,4)(82,7)(83,2)(84,7)(85,4)(86,5)(87,4)(88,2)(89,4)(90,3)(91,1)(92,3)(93,2)(94,4)(95,3)(96,5)(97,5)(98,3)(99,3)(100,2)(101,5)(102,6)(103,2)(104,3)(105,3)(106,6)(107,6)(108,1)(109,1)(110,5)(111,2)(112,3)(113,3)(114,4)(116,2)(117,1)(119,1)(120,3)(121,3)(122,1)(123,1)(125,1)(126,1)(127,2)(128,1)(131,1)(132,2)(133,1)(134,3)(135,1)(136,2)(137,1)(139,2)(144,1)(145,1)(146,1)(147,1)(151,1)(152,1)(154,1)(157,2)(159,2)(160,1)(164,2)(168,2)(169,1)(170,1)(178,1)(187,1)(190,1)(200,1)(203,3)(208,2)(211,1)(213,1)(214,1)(215,1)(216,1)(218,1)(219,2)(224,1)(231,1)(233,1)(237,1)(238,1)(239,1)(245,1)(248,1)(250,1)(252,1)(259,1)(265,1)(267,1)(275,1)(279,1)(288,1)(292,1)(297,1)(308,1)(310,1)(315,1)(337,1)(346,1)(352,1)(356,1)(383,1)(385,1)(393,1)(421,1)(422,1)(423,1)(441,1)(451,1)(492,1)(504,1)(520,1)(548,1)(579,1)(617,1)(643,1)(660,10)(668,1)(676,1)(701,1)(722,1)(723,1)(749,1)(756,1)(791,1)(801,1)(866,1)(915,1)(939,1)(948,1)(1015,1)(1232,1)(1252,1)(1301,1)(1592,1)(1750,1)(1772,1)(1782,1)(1946,1)(2049,1)(2255,1)(2750,1)(3308,1)(3722,1)(3809,1)(3827,1)(4886,1)(6234,1)(7879,1)(10782,1)(12778,1)(37916,1)
    };
    \legend{}
\end{loglogaxis}
\end{tikzpicture}

\vspace{0.25cm}

By restricting the graph to include only Romanian news sites (plus direct neighbors), we can still see a hint of the same pattern developing, but since the sample size is much smaller, we see outliers are more noticeable. In this graph we have 1404 nodes (of which 157 are news websites) and 2450 edges.

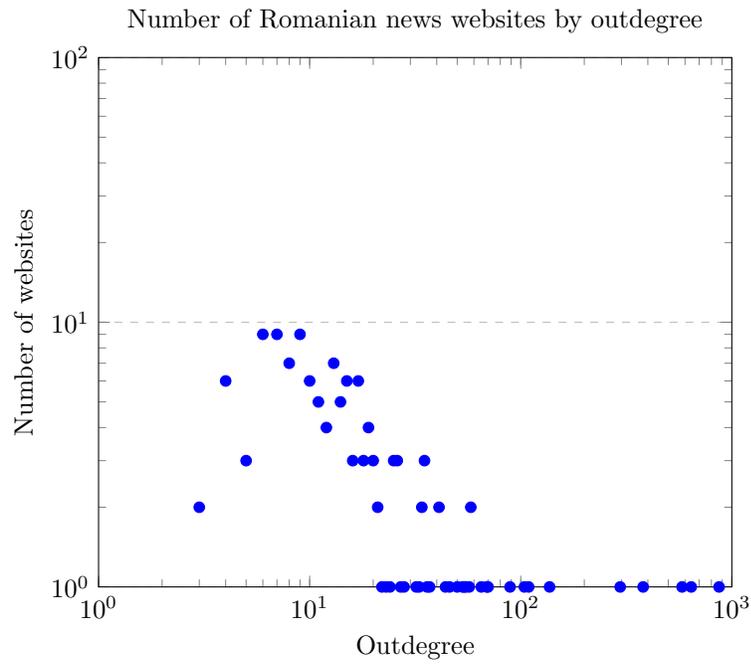
\vspace{0.25cm}

\begin{tikzpicture}
\begin{loglogaxis}[
    title={Number of Romanian news websites by outdegree},
    xlabel={Outdegree},
    ylabel={Number of websites},
    xmin=1, xmax=1000,
    ymin=1, ymax=100,
    xtick={1,10,100,1000},
    ytick={1,10,100},
    legend pos=north west,
    ymajorgrids=true,
    grid style=dashed,
]
\addplot[ color=blue, only marks ]
    coordinates {
    (0,5)(3,2)(4,6)(5,3)(6,9)(7,9)(8,7)(9,9)(10,6)(11,5)(12,4)(13,7)(14,5)(15,6)(16,3)(17,6)(18,3)(19,4)(20,3)(21,2)(22,1)(23,1)(24,1)(25,3)(26,3)(27,1)(28,1)(32,1)(33,1)(34,2)(35,3)(36,1)(37,1)(41,2)(44,1)(46,1)(50,1)(53,1)(54,1)(55,1)(57,1)(58,2)(65,1)(69,1)(70,1)(89,1)(104,1)(109,1)(137,1)(296,1)(380,1)(581,1)(642,1)(870,1)
    };
    \legend{}
\end{loglogaxis}
\end{tikzpicture}

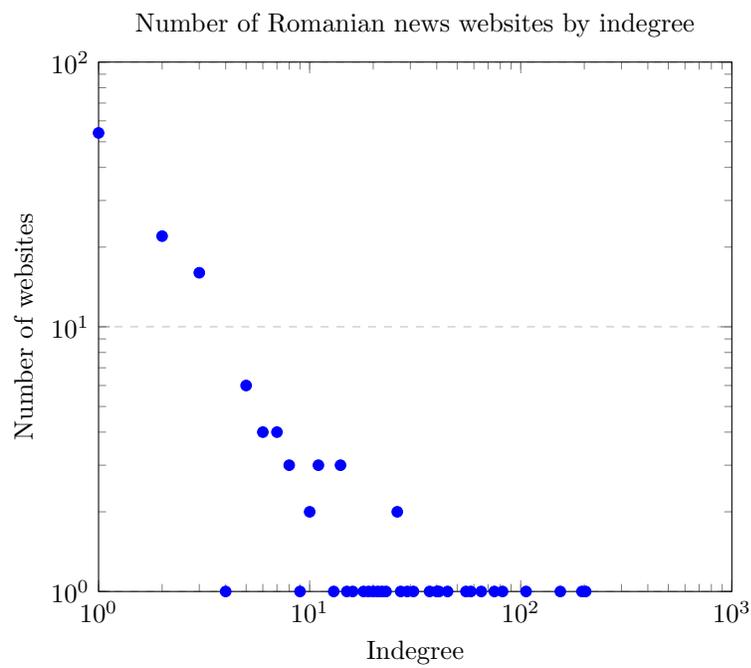
\vspace{0.25cm}

\begin{tikzpicture}
\begin{loglogaxis}[
    title={Number of Romanian news websites by indegree},
    xlabel={Indegree},
    ylabel={Number of websites},
    xmin=1, xmax=1000,
    ymin=1, ymax=100,
    xtick={1,10,100,1000},
    ytick={1,10,100},
    legend pos=north west,
    ymajorgrids=true,
    grid style=dashed,
]
\addplot[ color=blue, only marks ]
    coordinates {
    (1,54)(2,22)(3,16)(4,1)(5,6)(6,4)(7,4)(8,3)(9,1)(10,2)(11,3)(13,1)(14,3)(15,1)(16,1)(18,1)(19,1)(20,1)(21,1)(22,1)(23,1)(26,2)(27,1)(29,1)(31,1)(37,1)(40,1)(41,1)(45,1)(55,1)(58,1)(65,1)(75,1)(82,1)(106,1)(154,1)(195,1)(203,1)(1004,1)
    };
    \legend{}
\end{loglogaxis}
\end{tikzpicture}

The full data sets can be taken from CSV files stored on GitHub \cite{GitHubNewsCompareBackendResources}. This file format can be plugged straight into graph visualization software such as Gephi \cite{Gephi}, and potentially others with some tweaking. Since this graph of news websites effectively represents a social network, it exhibits all the standard properties of one, e.g., power law degree distribution, a high clustering coefficient, and a small diameter (relative to the number of nodes in the graph). We illustrate these properties below (the data points behind the plots are also available on GitHub \cite{GitHubNewsCompareBackendResources}).

\begin{table}[ht!]
\centering
\begin{tabularx}{\textwidth}{||X X X X X X||} 
 \hline
 Graph & Diameter & Average distance & Average degree & Clustering coefficient & Maximum degree \\ [0.5ex] 
 \hline\hline
 Total websites & 15 & 7.80 & 3.871 & 0.362 & 38198 \\
 News websites & 8 & 3.08 & 1.687 & 0.543 & 1874 \\
 \hline
\end{tabularx}
\caption{Romanian websites graph properties}
\label{table:GraphProperties}
\end{table}

It feels relevant to note that the website corresponding to the highest degree node (by an overwhelming margin) belongs to \textbf{hotnews.ro}, an online-only news outlet. We could interpret this both as a clearly focused effort to increase their footprint in the only arena where they are competing, but potentially also as a sort of underdog mentality, trying to make a disproportionate effort to get to the top and hold their position. It is likely that this strategy is paying off, considering how they are now considered one of the largest Romanian news websites, pulling in around 250.000-300.000 unique users daily and more than 3 million monthly unique visitors and around 30 million monthly page views, according to stats measured by the Romanian nonprofit organization BRAT (Romanian Joint Industry Committee for Print and Internet) \cite{SATIHotnewsRo2019}.

Other online news outlets that started out as more traditional media companies, like television
(e.g. \textbf{stirileprotv.ro} and \textbf{antena3.ro}), or print (e.g. \textbf{libertatea.ro}) appear to serve more as an extension of their main business, seeming to make little more than a token effort in establishing an online presence. The majority of outward links from these websites simply seem to promote other websites owned by the same parent company, while links from online-only outlets are a bit more varied and balanced.

\subsection{Social analysis based on the data}
\label{sec:news-article-analysis}

On the front of content comparisons, we have some interesting results showing similarities between distinct news outlets. We can see many instances where near-identical articles are displayed on different websites, with these websites sharing the same media group parent company (this is to be expected). If we filter out these cases, we then see instances where the application pinpoints articles about the same event, or on the same topic, with a fair rate of accuracy. While it would not be enough to conclusively pinpoint plagiarism, it is certainly a potential step in that direction, if we follow up on these leads (manually, for the time being). Gathering this kind of historic data can also be used to paint a picture about what kind of articles each particular outlet is liable to pick up on, and if we can notice any consistent groupings of websites emerging from there. A list of similar articles we have found over the course of running the application can be found in CSV format on Github \cite{GitHubNewsCompareBackendResources}.

To address fake news specifically, a recent report from Facebook \cite{FacebookBansPages} announces they have undergone efforts to remove what they call ``Coordinated Inauthentic Behavior'', i.e. pages that engage in manipulative behavior towards users on their platform. This is of particular interest to us, since some of these pages pose as Romanian news sources, which fits nicely into our use case. While the Facebook pages are no longer available, their associated websites are still alive and kicking: \textbf{destanga.ro, perele.ro, antifakenews.ro, momentulzero.ro}. These websites have not been discovered organically by our web crawler, despite having seen around 96,000 distinct URLs thus far, which would indicate that there are no links pointing to them at all in our entire data set. Out of curiosity, we add them to our list of target websites to see what we can learn from them, if anything. What we find is that they are largely isolated nodes in our graph, having very few distinct outward links, all of them pointing only towards Google, Facebook, or Wordpress.

Our text comparison component was only able to find very few matches involving these 4 websites over several runs, all of them  between \textbf{momentulzero.ro} and the ironically titled \textbf{antifakenews.ro}. This is a tentative indicator that at least some of these sites (labelled as misleading and manipulative by Facebook) are either coming from the same source or have the same goal in mind. Taking just a cursory glance at some of the articles served, we can see that they are quite short (around 500-1000 characters on average), and have no citations of any kind, even when alleging to use a direct quote from a particular person or institution (confirmed by our web crawler being unable to find any hyperlinks outside of social media). These are all good heuristic indicators that seem to support Facebook's conclusions in this particular case, and potentially lead us to other examples in need of a closer look.

\section{Conclusions and future work}

Throughout our inquiry, we manage to delve deep into the innards of our target websites, and glean some fairly intimate knowledge regarding their architecture and contents. Some of our expectations get challenged along the way, we might arrive at some surprising conclusions, and oftentimes the issues and challenges we come across can be particularly frustrating to get through, but still yield satisfying results. While we cannot expect groundbreaking results at every turn, or a ``smoking gun'' behind every corner, we trust that given enough time our application is capable of doing great things in capable hands. The amount of time saved by automating away cumbersome tasks empowers us to look at an increasingly larger picture, at a fine resolution. Sifting through this picture to find occasional nuggets of meaning can become a rewarding task in and of itself.

The current list of features and functionality included in the application is representative of the ideas we came up with, both on our own, and by studying existing research, all while timeboxing the implementation time to prevent ``feature creep'' so that the project does not drag on for many months or even years. We would be overjoyed to receive any kind of feedback from the community about our offering, and work with interested users to develop new features. We expect most of the future work involved will be around adding new statistics, reports and visualizations to the front-end, making it more friendly to people coming from a non-tech background. Barring some unforeseen revolutionary idea, the resulting data set gathered by the back-end component should be generic enough to be molded to match most reporting needs.

As mentioned by Marres and Weltevrede \cite{MarresN2012}, ``it would be a mistake to approach scrapers
as if they were stable, stand-alone machines: scrapers come in and fall out of use; they work, and then they no longer work''. We can certainly note that the stability of a particular piece of software is correlated with the amount of time spent bug-fixing, debugging and generally testing through use. To that end, making the app available to the public as a generic tool is probably the best way to find and fix the more glaring issues and omissions. After some growing pains, we expect to emerge with increasingly robust and battle-hardened versions of code, though some maintenance is likely to be required on a semi-regular basis, in case entirely breaking changes start to become widely adopted by target websites. To give a technical example, we can expect something like newly issued SSL certificates by certain certificate authorities to give us trouble if we are still using a particularly old version of the Java runtime that is unable to recognize them.

\bibliographystyle{abbrv}
\bibliography{bibliography}

\end{document}